\documentclass[%
reprint,
superscriptaddress,
showpacs,
 amsmath,
 amssymb,
 aps,
pra,
]{revtex4-1}
\usepackage{graphicx}
\usepackage{dcolumn}
\usepackage{bm}

\usepackage[colorinlistoftodos]{todonotes}
\usepackage{nicefrac}

\begin{document}

\preprint{APS/123-QED}

\title{Energy Harvesting with a Liquid-Metal Microfluidic Influence Machine}

\newcommand{\equalcontribution}{These authors have contributed equally to this work}
\newcommand{\CorrAuthor}{corresponding author: napoli@engr.ucsb.edu}

\author{Christopher Conner}\thanks{\equalcontribution}
\author{Tim de Visser}\thanks{\equalcontribution}
\author{Joshua Loessberg}\thanks{\equalcontribution}
\author{Sam Sherman}\thanks{\equalcontribution}
\author{Andrew Smith}\thanks{\equalcontribution}
\affiliation{Department of Mechanical Engineering, University of California Santa Barbara}
\author{Shuo Ma}
\affiliation{Department of Physics, University of California Santa Barbara}
\author{Maria Teresa Napoli}\thanks{\CorrAuthor}
\author{Sumita Pennathur}
\affiliation{Department of Mechanical Engineering, University of California Santa Barbara}
\author{David Weld}
\affiliation{Department of Physics, University of California Santa Barbara}

\begin{abstract}

We describe and demonstrate a new energy harvesting technology based on a microfluidic realization of a Wimshurst influence machine. The prototype device converts the mechanical energy of a pressure-driven flow into electrical energy, using a multiphase system composed of droplets of liquid mercury surrounded by insulating oil. Electrostatic induction between adjacent metal droplets drives charge through external electrode paths, resulting in continuous charge amplification and collection. We demonstrate a power output of 4 nW from the initial prototype and present calculations suggesting that straightforward device optimization could increase the power output by more than 3 orders of magnitude. At that level the power efficiency of this energy harvesting mechanism, limited by viscous dissipation, could exceed 90\%. The microfluidic context enables straightforward scaling and parallelization, as well as hydraulic matching to a variety of ambient mechanical energy sources such as human locomotion. 
\end{abstract}

\pacs{Valid PACS appear here}
\maketitle

\section{\label{sec:level1}Introduction}

Ambient mechanical energy from sources such as waves or human locomotion offers an appealing potential resource for new energy harvesting and portable power generation technologies~\cite{vibrationharvestingreview,harvestingreview,deQueiroz2013}. While recent progress in this area has been rapid, remaining limitations of many existing (e.g. piezoelectric and electromagnetic) mechanical harvesting technologies include fragility, low efficiency, and often the need for operation at or near a particular resonance frequency.
In this paper, we demonstrate a new technology for harvesting mechanical energy: a liquid-metal microfluidic Wimshurst machine, capable of attaining self-amplifying power output. Liquid metals have been incorporated in a number of energy harvesting technologies, such as devices based on reverse electrowetting~\cite{revelectrowetting1,revelectrowetting2}, thermo-pneumatic phenomena~\cite{thermopneumaticharvester}, triboelectric effects~\cite{triboelectricharvester}, and electret-droplet interactions~\cite{electretharvester}. Relatedly, non-metallic liquids have featured in promising microfluidic realizations of ballistic electrostatic generators~\cite{xie-ballistic} and Kelvin-type influence machines~\cite{microfluidickelvindropper,microfluidickelvindropper2}. We describe and present the first realization of a liquid-state Wimshurst machine, which directly transduces hydraulic power to DC electrical power without the use of bearings, magnets, surface electrochemistry, or piezoelectric elements. 

In general, the term ``influence machine'' refers to any generator which operates by electrostatic induction. Macroscopic influence machines based on rotating discs (as in the Wimshurst machine) or chains (as in the Pelletron) have been known since the nineteenth century, and are  commonly used to generate high voltages, for example in particle accelerators~\cite{PhysRev.24.690,influencemachines}. One of the most widely-known influence machines is the Wimshurst machine, which uses counter-rotating insulating disks with conducting patches.  Conducting brushes connect patches at opposite points of the same disk, enabling induction-based charge amplification, and collection electrodes harvest the charge for storage. 

\begin{figure*}[t!]
\centering
\includegraphics[height=0.36\linewidth]{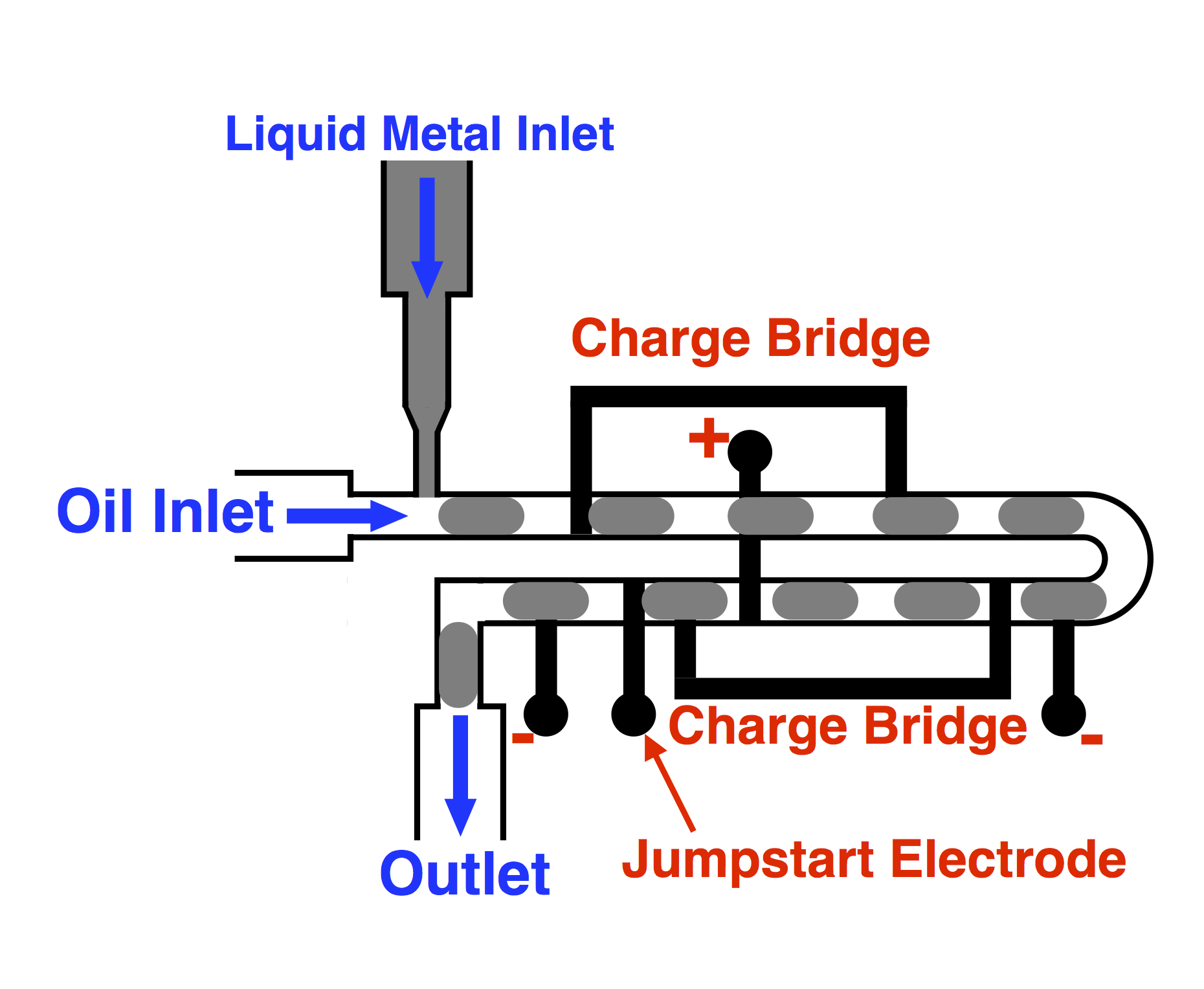}
\hspace{.6in}
\includegraphics[height=0.36\linewidth]{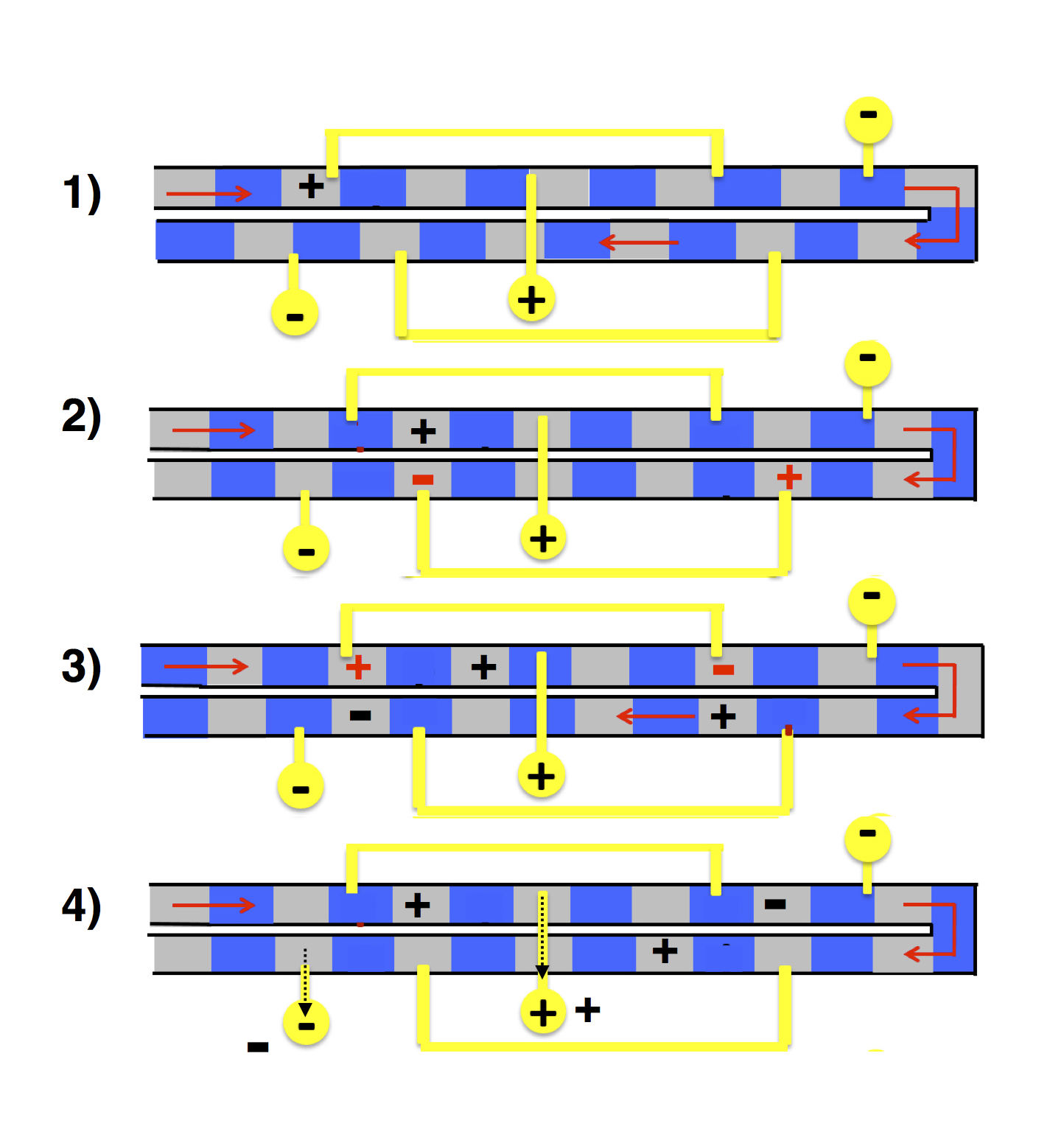}
\caption{\textbf{Left:} Schematic (not to scale) of liquid metal microfluidic energy harvester. Liquid metal droplets interact capacitively across the gap between channels, resulting in amplification of initial charge inhomogeneities. The ``jumpstart'' electrode can be used to set the sign of the initial charge. The electrodes labeled $+$ and $-$ collect the charge for external use or storage. \textbf{Right:}~Charge amplification by the Wimshurst mechanism. A liquid metal droplet shown in light gray starts with some inherent or seeded charge (1). When this charged droplet lines up with a droplet in the opposite channel which is connected to a ``charge bridge'' electrode (2), capacitive interactions across the gap between channels cause a charge transfer between the two liquid metal drops connected through the electrode. This process repeats as the metal droplets continue to flow (3). When the metal droplets encounter the collection electrodes, the excess charge is collected (4).}
\label{fig:schematic}
\end{figure*}

In the device we describe, shown schematically in Fig.~\ref{fig:schematic}, the rotating disks of a classical Wimshurst machine are replaced by parallel microfluidic channels containing oppositely-directed flows of alternating insulating and conducting liquid (oil and mercury, in the current prototype). The microfluidic context gives rise to intrinsic high efficiency, flexibility, reliability, and miniaturizability.  Hydraulic parallelization enables straightforward scaling to higher powers for a fixed input pressure and matching of harvester parameters to a variety of ambient mechanical energy sources.

Section~\ref{sec:concept} explains the quantitative theory of operation of this new energy harvesting device architecture, including a calculation of the maximum attainable harvesting efficiency.  Section~\ref{sec:fab} describes the physical design of our prototype device, and section~\ref{sec:results} presents the results of initial experiments demonstrating power generation.  Section~\ref{DevOpt} discusses pathways for optimization of device power and efficiency, and presents a concrete example of a possible energy harvesting application for this technology. Section~\ref{sec:conc} presents conclusions and some directions for future research.

\section{\label{sec:concept}Theory of Operation}

Fig.~\ref{fig:schematic} is a schematic depiction of the operating principle of our device: as alternating oil and metal droplets flow through a serpentine channel, capacitive interactions between adjacent conducting droplets drive charge through the external ``charge bridge'' electrodes, resulting in amplification of initial charge inhomogeneities. The charge is collected at the output terminals, which can be connected to a load for immediate use, or a capacitor for storage.   

To illustrate the promise of this new technology, we compute the theoretically achievable power, power density, and efficiency. The  theoretical maximum output power of such a device is straightforwardly calculated.  Just as in a macroscopic influence machine, the Wimshurst amplification mechanism should monotonically increase the charge per droplet up to a limit set by dielectric breakdown of the surrounding material (oil, glass, and PDMS in our design). 
For a given breakdown field $E_\mathrm{max}$, the surface charge density is limited to $\sigma_\mathrm{max}=\epsilon_0\epsilon_rE_\mathrm{max}.$
For physically reasonable values $\epsilon_r\simeq 5$ and $E_\mathrm{max}\simeq 10^7$~V/m,  $\sigma_\mathrm{max}$ is about 4$\times 10^{-4}$~$\mathrm{C}/\mathrm{m}^2$. In steady-state operation at maximum power output, a surface charge density of this order will be present on each metallic droplet passing the collection electrodes. Given a channel width $w$ of 300~$\mu$m and a flow velocity $v$ of 10~mm/s, easily achievable in our current prototype, the maximum average output current per channel for a device in steady-state operation is then
$I_\mathrm{max}= (4\pi/2)~\sigma_\mathrm{max}\cdot v\cdot w \simeq 8~\mathrm{nA}$.

Device operation near maximum surface charge density would result in
a droplet potential equal to
$
V_\mathrm{droplet} = q_\mathrm{max}/C_\mathrm{droplet} \simeq 1.5~\mathrm{kV}.
$
The maximum steady-state output power per channel is then
$
P_\mathrm{max}=V_\mathrm{droplet} \cdot I_\mathrm{max} \simeq 12~\mu\mathrm{W}.
$
This is a promising value, since it suggests that even without further optimization of the design, straightforward channel multiplexing should be easily capable of reaching the technologically relevant milliwatt regime. Since the maximum achievable output power is a critical figure of merit, it is important to identify which device and material parameters influence its value: 
\begin{align}
\label{eq:Pmax}
\begin{split}
P_\mathrm{max}&=\frac{q_\mathrm{max}}{C_\mathrm{droplet}}\cdot2\pi~\sigma_\mathrm{max}\cdot v \cdot w  \\
&=\epsilon_0 \epsilon_r E_\mathrm{max}^2 \pi~w^2~v.
\end{split}
\end{align}
This equation suggests a number of directions for optimization of device performance, as discussed in Section~\ref{DevOpt}.

There is another possible limit on device performance, due to droplet fission. According to the classic work by Rayleigh~\cite{rayleigh-fission}, the critical charge above which conducting droplets become electrostatically unstable is given by $\sqrt{64\gamma\pi^2\epsilon_0\epsilon_r R^3}$, where $\gamma$ is the interfacial tension and $R$ is droplet radius.  For our working fluids a conservative estimate of $\gamma$ is 0.35 N/m~\cite{Hg-interfacialtension}, giving rise to a maximum droplet charge around 200~pC.  This is actually a  less stringent limit than the breakdown limit discussed above, by about a factor of 2.  This estimate does not account for the field of the electrodes or the presence of confining channels which restrict the allowed droplet shapes; still, it provides an indication that the assumption of breakdown-limited operation is reasonably conservative.

The breakdown-limited maximum output power is produced by a single-channel device with an active volume $\Phi$ of a few cubic millimeters (device parameters are listed in Section \ref{sec:fab}). This device, which was not optimized for compactness, thus has a maximum volume power density $\Omega$ of
$
\Omega=P_\mathrm{max}/\Phi\simeq 5~\mathrm{mW/cm}^3.
$
This figure does not include the pump and capacitor. Scaling up to a larger number of channels will increase the percentage of the device volume occupied by the microfluidic charge amplifier, allowing the power density of the entire device to approach this bound.

The power efficiency of an energy harvester can be defined as the ratio of electrical output power to total input power. The efficiency of the microfluidic Wimshurst device will be limited by three terms: losses in the pump which generates flow, viscous dissipation in the microfluidic charge transport mechanism itself, and non-idealities in the storage and delivery of electrical power.  Pump losses will depend on the details of the pumping mechanism.  While our prototype uses a syringe pump to drive the flow, in an energy-harvesting context this could be replaced, for example, by a diaphragm pump integrated in a boot heel. Commercial diaphragm pumps can be up to 97\% efficient. Losses in power storage and delivery are well-understood and are common to essentially all forms of energy harvesting, so we will omit detailed discussion of them, except to say that the intrinsically DC nature of the power produced by this technology eliminates the need for lossy rectification circuits. The remaining term, viscous dissipation within the device, sets a fundamental upper bound for device efficiency.

\begin{figure}[t!]
\begin{center}
\vspace{-.15in}
\includegraphics[width=0.9 \columnwidth]{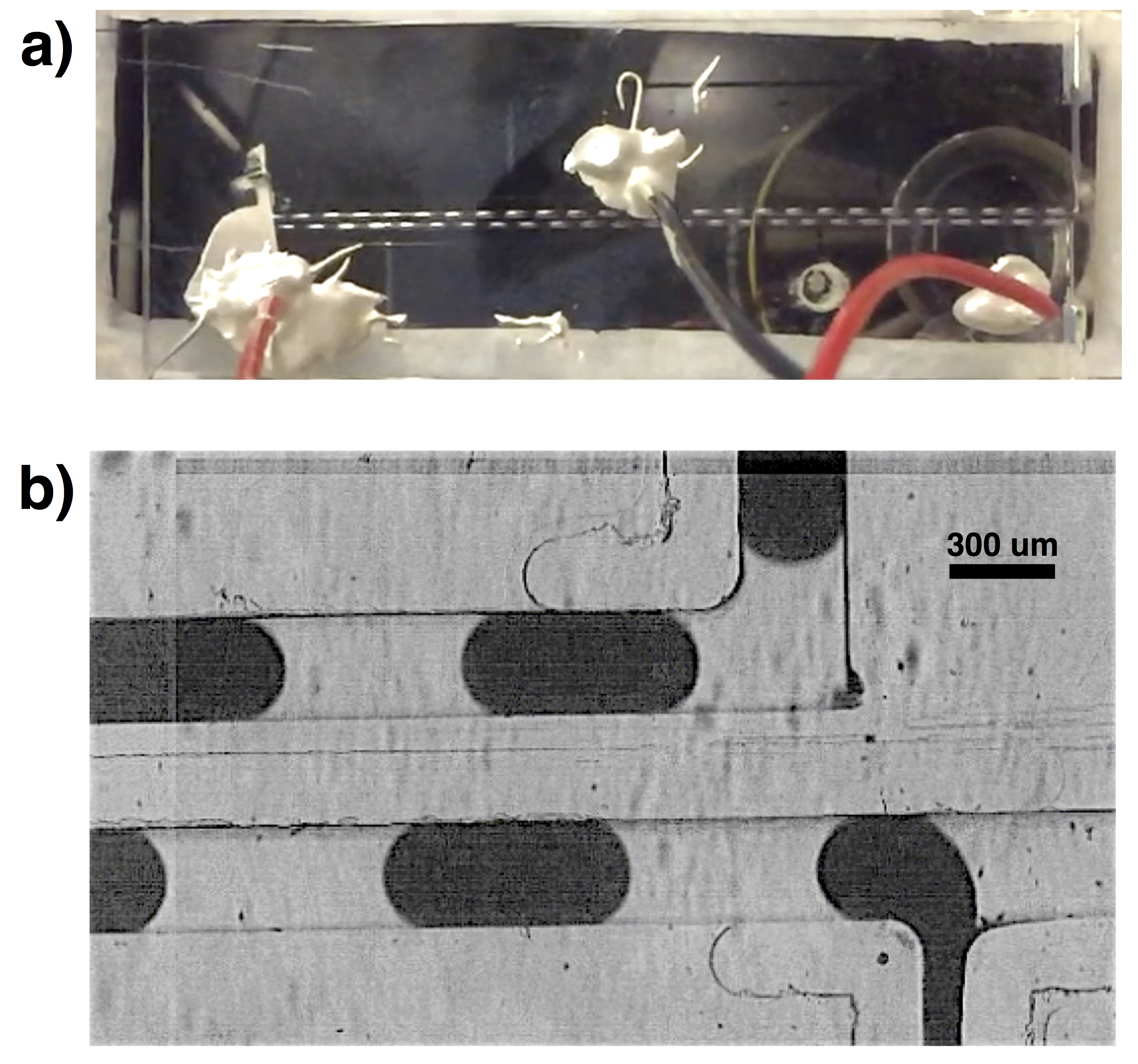}
\caption{\textbf{a:} Image of device during operation. \textbf{b:} Photograph of liquid metal droplets being generated and moving through prototype device.}
\label{fig:blobs}
\end{center}\end{figure}

The pressure drop due to viscous drag in the microfluidic channels is given by the Darcy-Weisbach relation:
\begin{equation}
\Delta P = \rho f \frac{Lv^2}{2w} = 32~\eta \frac{L~v}{w^2}
\label{DWrelat}
\end{equation}
where $\rho$ is the mass density of the fluid, $f\equiv 64/\mathrm{Re}$ is the Darcy friction factor, $\mathrm{Re}\equiv \rho~v~w/\eta$ is the Reynolds number, $L$ is the channel length, $\eta$ is the viscosity, and $v$ and $w$ are, as above, the velocity of the flow and width of the channel. Since the volume flow rate $Q$ through the channel is simply $Q=v~w^2$, the dissipated power due to viscous drag is
\begin{equation}
P_{dissipated}=Q \cdot \Delta P = 32 \eta~L~v^2.
\label{eq:disspower}
\end{equation}
Assuming viscous drag is the dominant dissipation mechanism, the maximum power efficiency $\alpha$ is thus the ratio between maximum electrical output power (given in Eq.~\ref{eq:Pmax}) and total input power (given by the sum of Eqs.~\ref{eq:Pmax} and \ref{eq:disspower}):
\begin{equation}
\alpha=\frac{P_\mathrm{max}}{P_\mathrm{max}+P_\mathrm{dissipated}}=\frac{\epsilon_0 \epsilon_r E_\mathrm{max}^2 \pi~w^2~v}{\epsilon_0 \epsilon_r E_\mathrm{max}^2 \pi~w^2~v + 32 \eta~L~v^2}.
\end{equation}
For our device parameters (see Section \ref{sec:fab} below), and physical values for viscosity (about 1.5 cP for both mercury and perfluorooctane), this maximum efficiency is 98\% for a flow velocity of 10 mm/s, and 90\% for a flow velocity of 50 mm/s.  This number does not include losses in the pump or the charge storage mechanism, and it assumes device operation at the maximum electrical output power given by Eq.~\ref{eq:Pmax}. Additionally, in applying the Darcy-Weisbach relation we have implicitly assumed that viscous drag dominates over other dissipation mechanisms such as wall friction; this is an assumption that can be tested experimentally in future devices. Still, this is clearly a promising upper bound when compared to existing forms of energy harvesting, and this result suggests that viscous dissipation can be easily dominated by power generation for achievable device parameters.

\section{\label{sec:fab}Device Design and Operation}

The prototype device (see schematic in Fig.~\ref{fig:schematic}) 
was constructed using standard microfluidic fabrication techniques, discussed in detail in appendix~\ref{fab}. Briefly, a polydimethylsiloxane (PDMS) channel geometry is molded from an aluminum mold and bonded to a micromachined fused silica substrate with lithographically patterned electrodes. All channels are 300~$\mu$m deep and 300~$\mu$m wide, except for the 130-$\mu$m-wide inlet channel for the liquid metal. The separation between the two sections of the serpentine channel is also 300~$\mu$m. The main channel is 5~cm long from the T-junction to the bend. Access ports in the PDMS layer are 2~mm in diameter to allow for hydraulic and electrical connections. An image of our completed device as well as a photograph of the liquid metal/oil droplet flow within the serpentine channel is shown in Fig.~\ref{fig:blobs}a.

\begin{figure}[t!]
\centering
\includegraphics[width=0.9\linewidth]{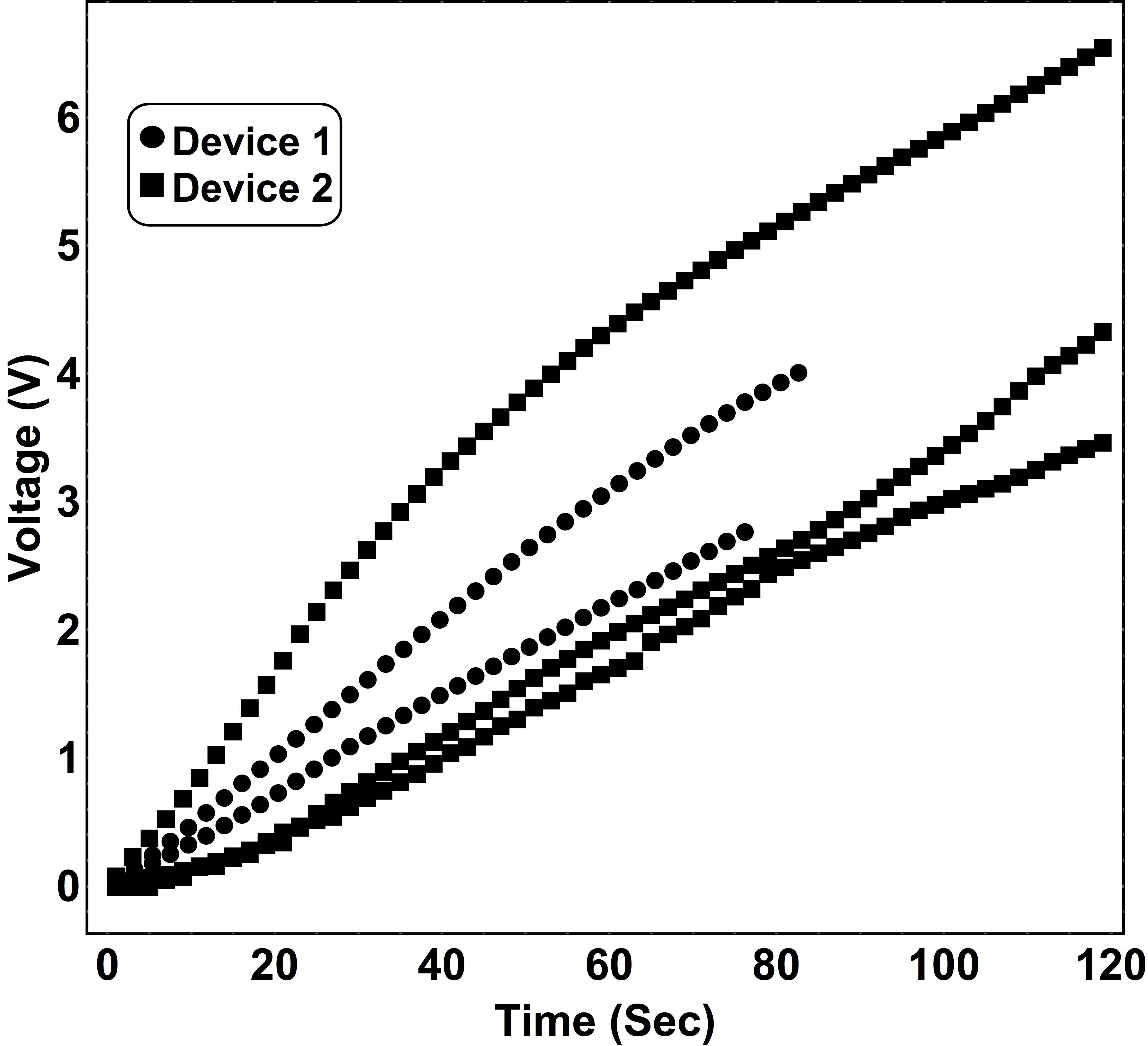}
\caption{Absolute value of output voltage versus time for two separate tests with two separate devices. Voltage is measured at the 10~nF collection capacitor.}
\label{fig:sampledata}
\end{figure}

Liquid metal droplets are generated by the microfluidic junction shown in Figure \ref{fig:schematic} and visible in Fig.~\ref{fig:blobs}b. Charge is collected by a 10 $\mathrm{nF}$ capacitor connected to the output electrodes. To facilitate droplet production, the mercury inlet at the T-junction is narrower than the oil inlet (130~$\mu$m). The consistent generation of regularly spaced metal droplets of length $L_{\mathrm{metal}}$, roughly equal to their separation $L_\mathrm{oil}$, depends upon appropriate inlet flow rates and the use of relatively long metal droplets (aspect ratio~$\simeq$2). We find that long metal droplets perform better than shorter ones, because the latter end up being encapsulated by oil, which in turn changes the droplet spacing during flow and can disrupt charge transfer. The flow rates were selected to obtain the fastest possible flow consistent with droplet regularity, while maintaining roughly constant $L_\mathrm{oil}\simeq L_{\mathrm{metal}}$. For our current prototype dimensions, these flow rates were set to 30 ml/hr for the oil and 15 ml/hr for the metal, which resulted in generation of droplets about 700 $\mu \mathrm{m}$ in length at a frequency of about 86 droplets per second. Flow rates much greater than these values were observed to lead to increased variance in droplet size.

\section{\label{sec:results}Results}

Fig. \ref{fig:sampledata} shows experimental data from two separate devices, which demonstrates successful energy conversion using our prototype. Specifically, all experimental trials consistently generate voltage across the connected 10 nF capacitor, with roughly linear growth.

\begin{figure}[t]
\centering
\includegraphics[width=0.8\linewidth]{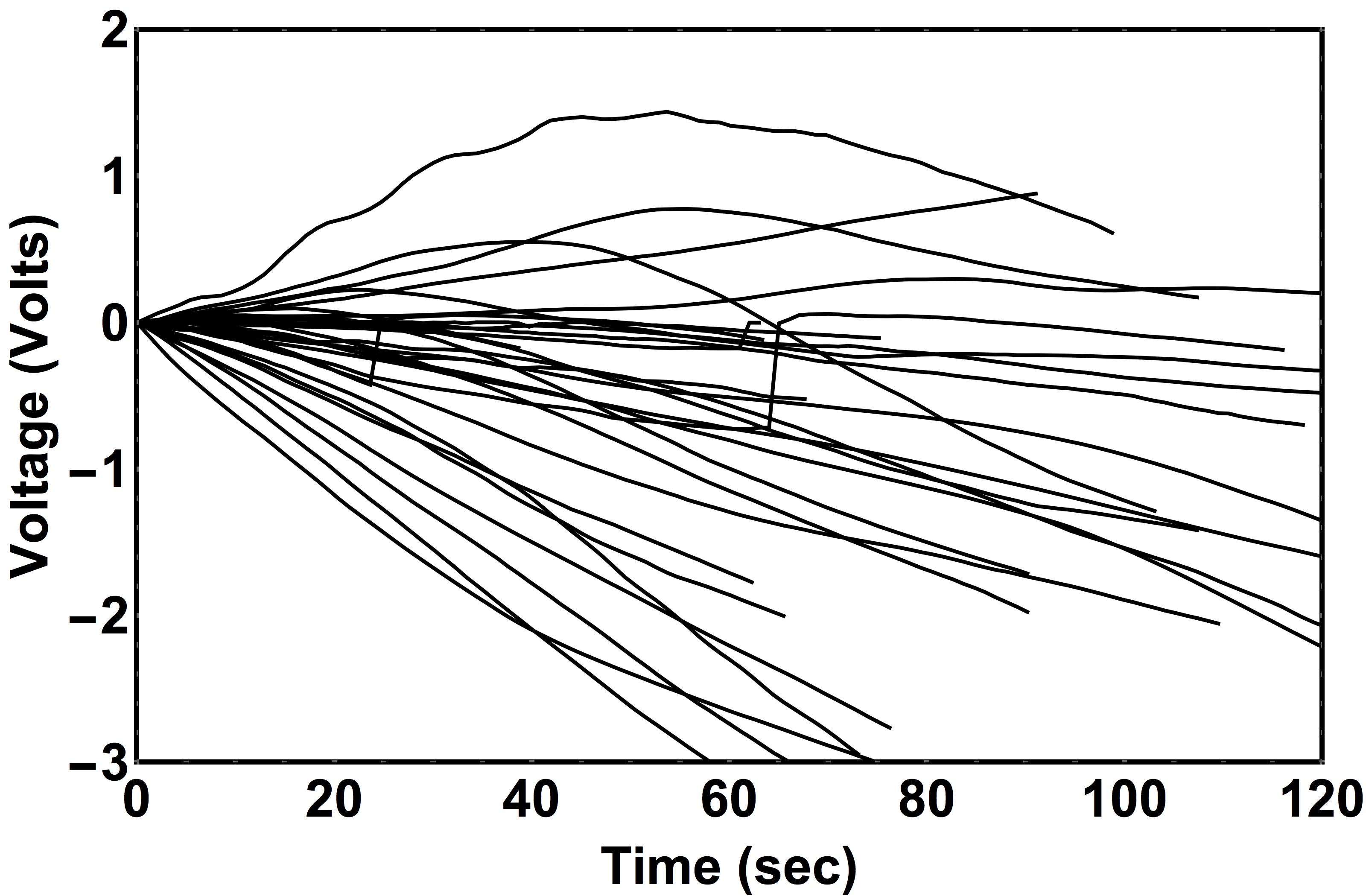}
\caption{Voltage versus time for all data runs  with no jumpstart voltage from Fig.~\ref{fig:bargraph}. Note the variability in direction of voltage growth.}
\label{fig:variability}
\end{figure}

\begin{figure}[t]
\centering
\includegraphics[width=0.9\linewidth]{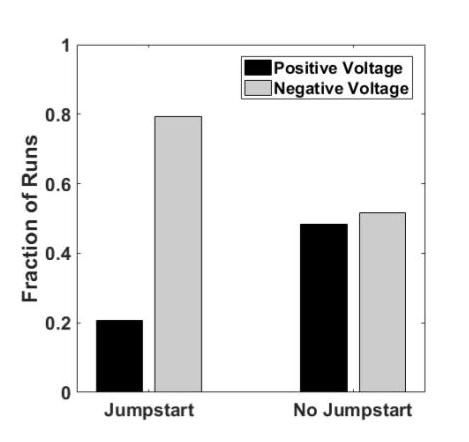}
\caption{Comparison of the direction of voltage growth in data runs with and without a jumpstart applied. }
\label{fig:bargraph}
\end{figure}

The sign of the voltage growth, normally induced by random fluctuations of droplet charge, can be controlled by briefly applying a voltage to the ``jumpstart'' electrode (see Figure \ref{fig:schematic}), thereby forcing a known polarity on the inflowing droplets. To demonstrate this effect, we performed 58 experiments with a jumpstart voltage of $+10$ V, applied during the first 3 seconds of the experiment, and 31 runs without a jumpstart voltage. With no jumpstart applied, there was an approximately equal chance of generating positive or negative voltage: 15 of 31 test runs had a positive voltage and 16 of 31 had a negative voltage (Fig. \ref{fig:variability}). With an applied jumpstart, this distribution shifted radically in favor of negative voltage: 80\% (46 of 58) of experiments produced negative voltage (Fig. \ref{fig:bargraph}). Experiments without a jumpstart voltage also occasionally exhibited a change in the sign of voltage growth during the run (Fig. \ref{fig:variability}). Application of the jumpstart voltage largely eliminated this phenomenon. 

Our experimental results show that the power produced by the prototype device is substantially smaller than its theoretical maximum. Experiments lasting a few minutes did not approach the breakdown-voltage limits of the device, where maximum power generation is expected. For the data shown, the maximum potential difference produced was 6 V over two minutes, and the voltage growth rate varied from $25\ \mathrm{mV}/\mathrm{sec}$ to $67\ \mathrm{mV}/\mathrm{sec}$. This corresponds to a maximum power output of 4 nW and an efficiency (based on the assumption of viscous loss described in Eq.~\ref{DWrelat}) of less than 2\%; neither number is yet close to the theoretical maximum for the device geometry. Additionally, there is some variation in rates of voltage generation even with an applied jumpstart voltage,  and growth behavior is linear, not exponential. Several factors may limit the power and contribute to power variability, including fluctuations in droplet size, inefficient charge transfer across the thin oxide barrier which protects the electrodes, and subcritical capacitive coupling between droplets in adjacent channels. 

\begin{figure}[t]
\centering
\includegraphics[width=0.95\linewidth]{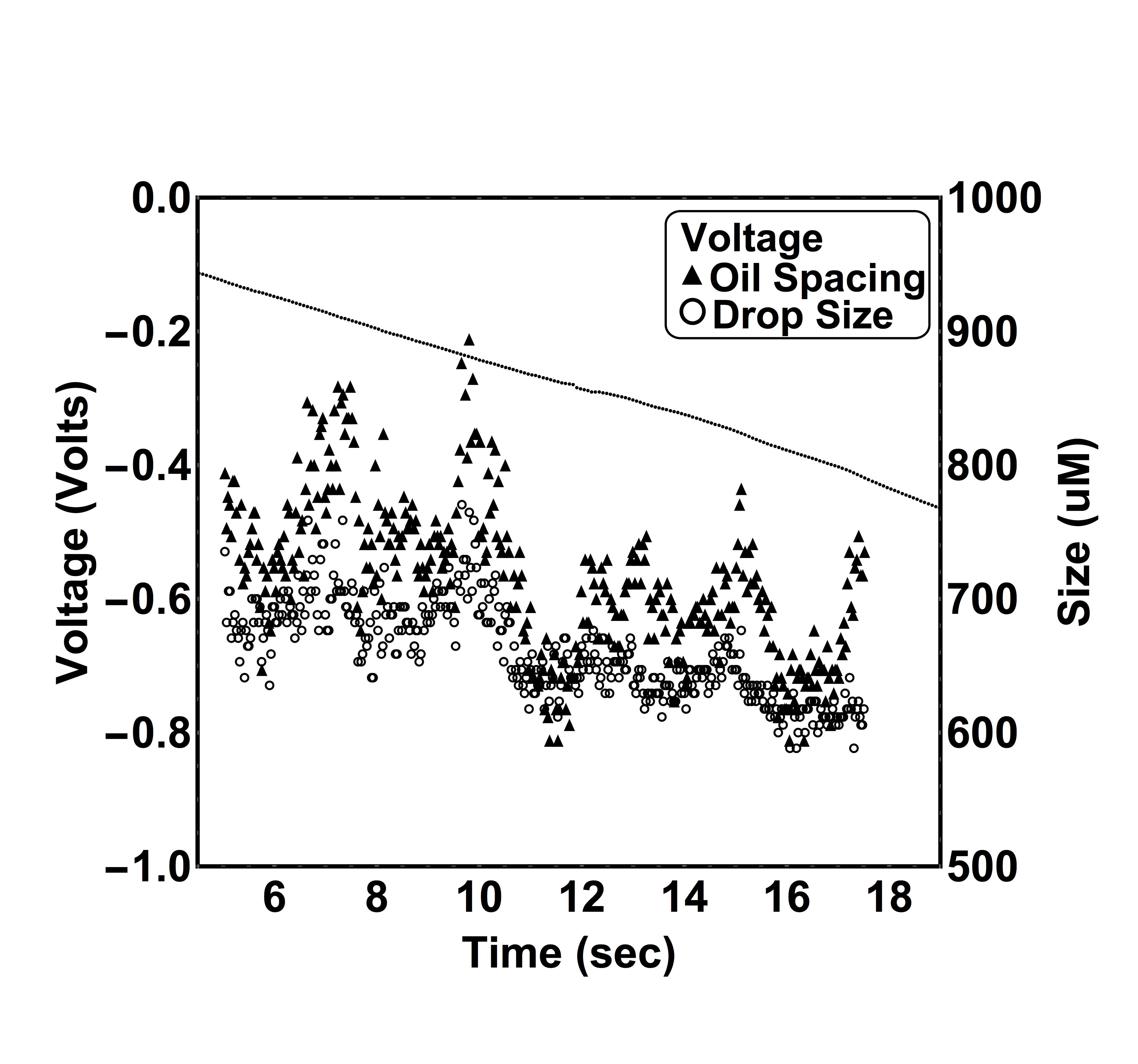}
\caption{Droplet sizes and voltages versus time during a test of the prototype device.}
\label{fig:droplets}
\end{figure}

To investigate the first possibility (droplet size variations), we measured the voltage growth versus time, while simultaneously measuring the droplet sizes with video analysis. Fig~\ref{fig:droplets} shows the measured voltage and measured droplet sizes over a period of 15 seconds. From these data, it is evident that the droplet size can vary by up to 150 $\mu$m without substantially affecting the instantaneous voltage growth rate. 
Another related possible reason for inconsistencies in the voltage growth rate is changes in the relative spatial phase of the droplets and the brush electrodes. For proper functioning of the device, droplets must line up over both ends of the brush electrodes simultaneously so that charge can transfer; in principle, droplet spacing variations can violate this condition. However, analysis of video data confirmed that nearly every droplet makes a connection with another droplet each time it passes a bridge electrode, even in the presence of droplet size variations larger than those shown in Fig.~\ref{fig:droplets}. Therefore, droplet size variability and spatial phase drift do not appear to be responsible for limiting the output power of the device. 

Another potential reason for power variation is poor charge transfer through the oxide layer covering the electrodes. A straightforward test of this was achieved by filling the entire channel with liquid metal and measuring the resistance between pairs of electrodes. The resistance was negligible, suggesting that charge transfer from the liquid metal to the electrodes is not an important limitation on the power produced by the prototype device.

\begin{figure}
\centering
\includegraphics[width=0.55\linewidth]{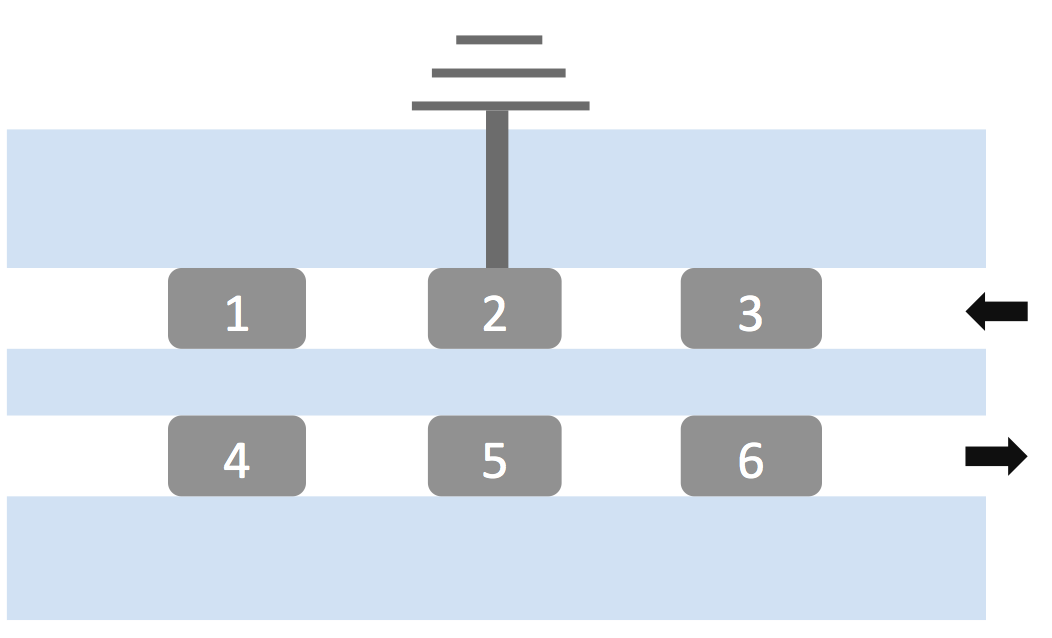}

\vspace{.2in}

\includegraphics[width=0.55\linewidth]{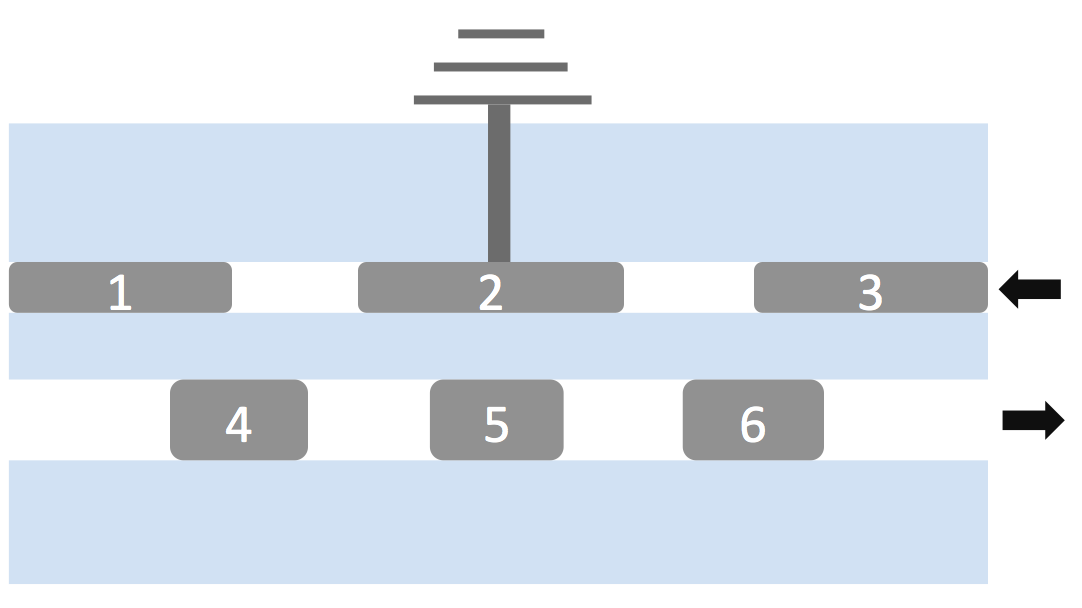}
\caption{Droplet geometry used for calculating Wimshurst amplification mechanism.  Discussion of the calculation  in the text refers to the droplet labels shown here. \textbf{Top:} Standard-width channel layout, as in prototype device.  \textbf{Bottom:} Channel layout with reduced width near charge bridge electrode enhances capacitive coupling between the droplet being charged and droplets in the opposite channel.}
\label{fig:capacitance}
\end{figure}

Finally, the rate of voltage growth in the prototype device may be limited by insufficient capacitive coupling between the two channels. The Wimshurst amplification mechanism sketched in Fig.~\ref{fig:schematic} leads to exponential voltage growth only if the charge induced on the droplet connected to the bridge electrode is larger than the charge per droplet in the opposite channel.  This condition can be satisfied or not, depending upon the details of the geometrical capacitive coupling between droplets.  A simple estimate of the charge amplification factor can be obtained by considering the droplet arrangement shown in Fig.~\ref{fig:capacitance}.  If the charge and voltage on the six droplets are represented as $\vec{Q}$ and $\vec{V}$ respectively, so that $Q_i$ is the charge on the $i$th droplet, then the system satisfies the set of six linear equations $\vec{Q}=\mathbf{C}\vec{V}$, where $\mathbf{C}$ is the Maxwell capacitance matrix. We have estimated the elements of $\mathbf{C}$ for our device geometry both numerically and analytically. In the arrangement shown in Fig.~\ref{fig:capacitance}, the droplets in the upper channel are passing by a bridge electrode, approximated here as an electrical ground.  We make the following assumptions: droplets 1 and 2 have been charged to a final charge $q_f$ by the electrode, droplet 3 has negligible charge as it has just come from the collection capacitor, and droplets 4, 5, and 6 all have identical charge $q_i$ (which is the opposite polarity from $q_f$).  All droplet potentials are free parameters except for the potential of droplet 2, which is enforced to be zero by the electrode. The Wimshurst amplification factor $\Gamma \equiv -q_f/q_i$ can be calculated by simultaneously solving the six equations under these assumptions.  For the existing device geometry, such a calculation indicates that $\Gamma \simeq 0.4$. This supports our hypothesis that geometrical effects are limiting the rate of voltage growth in the prototype and leading to the observed linear (as opposed to exponential) voltage growth in time. 

\section{\label{DevOpt}Device Optimization}
The microfluidic context enables a simple and powerful method for increasing $\Gamma$ which is impossible in solid-state influence machines: variation of the channel width and separation.  For example, if the width of the upper channel in Fig.~\ref{fig:capacitance} is reduced by a factor of 2, the incompressible metal droplets in that channel would double their length (Fig.~\ref{fig:capacitance}b).  It is clear from geometrical considerations that $C_{24}$, $C_{25}$, and $C_{26}$ would then be enhanced relative to $C_{12}$ and $C_{23}$.  Physically, this means that the oppositely-charged droplets in the bottom channel would induce a larger charge on the droplet contacting the electrode, and the adverse effect of the same-polarity droplet charges in the upper channel would be reduced. Reducing the separation between channels should also enhance the desirable capacitive couplings.  These intuitive expectations can be easily checked using the Maxwell capacitance matrix formalism and numerical calculation of interdroplet capacitances. The results of such a calculation indicate that an upper-channel width of 100~$\mu$m and a channel separation of 50~$\mu$m, with all other parameters the same as in the current prototype, would lead to values of $\Gamma$ greater than 1.1. This demonstrates that morphological optimization is a promising and unique direction for future development of microfluidic influence machines. Straightforward improvements to the device geometry should allow us to enter the region of the parameter space characterized by exponential voltage growth, thereby substantially increasing the device power.

It is also possible to increase the output power using several other simple improvements. For example, from Eq.~\ref{eq:Pmax}, the output power can be increased quadratically by increasing the channel width $w$ and linearly by increasing the flow velocity $v$. In addition, the indicated dependence of output power on dielectric and breakdown properties of the non-conducting components of the device opens up further opportunities for optimization. 

The microfluidic context also enables scalability via multiplexing and large-scale integration. Individual single-channel devices can be combined together in series or in parallel both hydraulically and electrically. In fact, hydraulic and electrical connection architectures are independent of one another, enabling matching of an up-scaled device to particular energy source and load characteristics. A 1000-channel device can produce useful power in the range of 10 mW. 

As a concrete example of energy-harvesting possibilities for this technology, we briefly consider whether such a scaled-up device can be conveniently powered by a specific ambient energy source: human locomotion.  For concreteness, we consider a 10 mW harvesting device  powered by a diaphragm pump in a boot heel.  Ground pressure from an average-sized walking human is about 50-60 kPa.  From the efficiency calculations above, it is clear that the pressure drop across a single channel is dominated by the electrostatic back pressure of around 14 kPa at maximum power.  This suggests a configuration of 250 parallel hydraulic channels each of which contains 4 amplifier segments in series, for a total pressure drop of 56 kPa, well-matched to the pressure produced by a human foot. The total volume flow rate through all channels in such a device, assuming individual device parameters matching those of our prototype, would be about $2\times10^{-7}$ m$^3$/s. Such a flow could be produced by a 2-cm-diameter diaphragm pump with a heel displacement of half a millimeter, assuming one step per second. 
There is thus good reason to expect that a scaled-up version of our prototype device could be portable, practical, and sufficiently powerful for a variety of energy harvesting uses.

\section{\label{sec:conc}Conclusions and Outlook}
We have proposed and demonstrated a microfluidic energy harvester based on a liquid-state realization of a Wimshurst influence machine. Calculations indicate that straightforward improvements to the geometry should be capable of increasing the output power of a single-channel device by up to three orders of magnitude. Additional future improvements to the device include its realization in glass rather than PDMS (to increase robustness) and the use of a non-toxic liquid metal such as GaInSn, or perhaps a liquid semiconductor. The high intrinsic efficiency of the technique, as well as the scalability and parallelizability inherent in the microfluidic format, give this new technology substantial near-term promise for real-world energy harvesting applications. 

\begin{acknowledgments}
The authors thank A.~C.~M.~de~Queiroz for a useful and interesting discussion. This material is based upon work supported in part by the National Aeronautics Space Administration under Contract No. NNX16CS10P and NNX14CS61C awarded to Angstroms Designs, Inc. Additionally, this work was partially supported by the Institute for Collaborative Biotechnologies through grants W911NF-09-0001 and W911NF-12-1-0031 from the U.S. Army Research Office. The content of the information does not necessarily reflect the position or policy of the U.S. Government, and no official endorsement should be inferred. Finally, we acknowledge awards from the UCSB Academic Senate (\# 565020-19941) and summer intern programs through the National Science Foundation Grant No. 1402736.
\end{acknowledgments}

\bibliographystyle{unsrt} 

\appendix
\section{\label{fab}Fabrication}
The LIMMPET device consists of a PDMS structural layer, containing the microfluidic channel and access ports, connected to a silica substrate with embedded electrodes. 

The microfluidic channel is obtained from an aluminum mold, which reproduces the channel geometry of Figure \ref{fig:schematic}. Except for the mercury inlet (130~$\mu$m wide), all channels are 300~$\mu$m wide. All channels (including the mercury inlet) are 300~$\mu$m deep. A thin layer of canola oil is applied to the aluminum mold before each use to help the PDMS release from the mold after baking. Approximately 10~g of PDMS (10:1 base to curing agent) is poured into the mold, which is then placed in a vacuum chamber (Fisher Scientific MaximaDry) for 20~minutes to eliminate any bubbles, and then baked at 100$^\circ$C for 45~minutes in an oven (Yamato DKN400). The baked PDMS is then removed from the mold, wiped down with acetone, rinsed with DI water, and dried with a nitrogen gun.

The PDMS microfluidic device is then bonded to a micromachined fused silica substrate (HOYA) which contains embedded electrodes, as schematically shown in Figure \ref{fig:schematic}.  Using one lithographic step, 100-nm-deep trenches are wet etched into the substrate, using a buffered oxide etchant (HF:H$_2$O in a ratio 1:6), then 100~nm of titanium is deposited across the wafer via e-beam evaporation. The trenches, which are filled with titanium, are then patterned via a lift-off method using acetone and isopropanol soaking steps. The substrate is then exposed to 2~minutes of O$_2$ plasma at a pressure of 300~mTorr and a power of 100~W to descum the surface. The final step of the micromachining process involves depositing 10~nm of silicon dioxide  across the surface of the wafer via plasma enhanced chemical vapor deposition (PECVD). A profilometer (Dektak 6M) scan of the surface reveals a very planar surface, with trenches due to the embedded titanium traces measuring no more  than 7~nm across the wafer. 

Next, the PDMS layer is bonded to the silica substrate. Both the PDMS layer and silica substrate are ozone-treated for 10~minutes (Novascan PSD Pro Series Digital UV Ozone System), then aligned and placed into contact. Light pressure is applied to force air bubbles out from between the PDMS and substrate. The device is then baked for 30~minutes at 100$^\circ$C on a hot plate. After bonding, 2-mm-diameter holes for the collector and jumpstart electrodes are cut using a laser cutter. Careful control of the power and time of the laser cutter recipe allows for only the PDMS to be removed. Finally, the inlet and outlet tubes are inserted and held in place with epoxy. The wires for the electrodes are attached with conductive epoxy.

\end{document}